\documentclass[12pt,a4paper]{article} 
\usepackage{epsfig}
\usepackage{graphics}
\usepackage{subfigure}
\usepackage{a4wide}
\usepackage{amsmath}

\title{Skyrmions on the Two-Sphere.}
 \author{M de Innocentis\footnote{email: marco.de-innocentis@durham.ac.uk}
 and R S Ward\footnote{email: richard.ward@durham.ac.uk}
 \bigskip
\\Department of Mathematical Sciences,  \\ University of
Durham, \\Durham DH1 3LE}

\newcommand{\pa}{\partial}

\begin{document}
\maketitle \abstract{\noindent
We study static solutions of the Skyrme model on the two-sphere
of radius $L$, for various choices of potential.  The
high-density Skyrmion phase corresponds to the ratio
$\beta=L/{\rm(size\ of\ Skyrmion)}$ being small, whereas the
low-density phase corresponds to $\beta$ being large.  The transition
between these two phases, and in particular the behaviour of
a relevant order parameter, is examined.}

\section{Introduction}

Topological solitons are usually studied on flat space, but there are
several reasons why the curved-space setting is interesting (apart from
the most obvious one of possible cosmological applications).  First,
it often reveals useful mathematical features; there are now two length
scales, namely the size of the solitons and the radius of curvature of the
underlying geometry, and there is an interplay between these scales.  For
monopole-like objects such as as vortices and Yang-Mills-Higgs monopoles,
one can work on hyperbolic space: in the case of vortices, the equations
for static vortices become integrable for a particular value of the
curvature \cite{Wit77}, \cite{Str92}; and for YMH monopoles,
the system is simpler on hyperbolic space than on flat space \cite{At84}.
For textures, such as the Skyrmions discussed in this paper, one may
study the problem on either hyperbolic spaces or spheres ({\it ie} on spaces
of constant negative or positive curvature), and this similarly throws up
interesting geometrical features \cite{MR86}, \cite{Man87}, \cite{JMW89},
\cite{MR90}, \cite{SB98}, \cite{H98}, \cite{Kr00}.

There are also more physical reasons for studying textures on compact
spaces, namely that this models a finite density of solitons, and the
transistion between the high-density and the low-density phases.  One may do
this without introducing curvature, by taking space to be a flat torus;
but it is in some ways simpler to use a sphere.
This was first done for the Skyrme model in three spatial
dimensions, which is a model of nuclear matter.  But there
has also been interest in the two-dimensional version, both
as a `toy' model and also because it has several potential
applications in condensed-matter physics.  In this paper, we
study the Skyrme model on the two-sphere of radius $L$, and in
particular the dependence on the ratio
$\beta=L/{\rm(size\ of\ Skyrmion)}$.

In the three-dimensional Skyrme model (without a pion mass term), the
Skyrmion for small $\beta$ is homogeneously spread out over space; when
$\beta$ reaches a certain critical value, the Skyrmion localizes around
a point \cite{Man87}, \cite{Kr00}.  If the winding number is greater than
one, or if a zeroth-order potential term (such as a mass term) is present,
then this transistion is less clear-cut.  In the two-dimensional
case, such a potential term is obligatory, and so this smearing effect is
inevitable ({\it cf}\ \cite{SB98}).  The quantity $\Phi$ which
monitors the deviation from homogeneity tends to zero as $\beta$ does, but
$\Phi$ is never quite zero, and so is not an order parameter in the strict
sense.  There is, however, another quantity $\Psi$ which does serve as an
order parameter, namely one which monitors the breakdown of a certain
reflection symmetry \cite {JMW89}, \cite{Kr00}; and the associated phase
transition is sharp (as long as the potential
has this symmetry, which was not the case for the system studied in
\cite{SB98}).  We shall investigate these phenomena, for several different
potentials, by finding rotationally-symmetric solutions numerically.

\section{Skyrme models on the two-sphere}

In this section, we review the Skyrme system on the two-sphere, the
topological Bogomol'nyi bound on its energy, and the imposition of
rotational symmetry.  Let $M$ denote the standard 2-sphere of radius $L$,
with local coordinates $(x^1,x^2)$ and metric $ds^2 = g_{jk}\,dx^j\,dx^k$.
The area element on $M$ is $\omega=h\,dx^1\wedge dx^2$, where
$h=[\det(g_{jk})]^{1/2}$.  The Skyrme field $\phi$ is a map from $M$ to
the unit sphere $S^2$, and may be thought of as a unit vector field
$\vec\phi(x^j)$ (so $\vec\phi\cdot\vec\phi=1$).  We are considering only
the static problem, so  $\vec\phi$ depends only on the spatial variables
$x^j$, and not on time.  Let $\pa_j\vec\phi$ denote the partial derivative
$\pa\vec\phi/\pa x^j$.
The winding number (degree) of $\vec\phi$ is denoted by the integer $k$.
If we define the function $\rho[\vec\phi]$ as the triple scalar product
$\rho[\vec\phi] = 2\vec\phi\cdot(\pa_1\vec\phi)\times(\pa_2\vec\phi)$, then
(assuming that the $x^j$ cover almost all of $M$) we have
\begin{equation} \label{k}
 k = \frac{1}{8\pi}\int_{M} \rho[\vec\phi]\,dx^1\,dx^2.
\end{equation}
The (normalized) energy $E_k$ of $\vec\phi$ is taken to be a sum of three
terms $E_k = E^{(2)} + E^{(4)} + E^{(0)}$, with each being the integral over
$S^2$ of the corresponding density function:
\begin{equation} \label{Ek}
 E^{(n)} = \frac{1}{4\pi |k|}\int_{M} {\cal E}^{(n)}[\vec\phi]\, \omega
\end{equation}
for $n=2,4,0$.  The densities  ${\cal E}^{(n)}(\vec\phi)$ are given by
\begin{eqnarray}
 {\cal E}^{(2)}[\vec\phi] &= &\frac{1}{2}g^{jk}(\pa_j\vec\phi)
           (\pa_k\vec\phi),   \label{E2} \\
  {\cal E}^{(4)}[\vec\phi] &= &\frac{1}{8}\gamma h^{-2}\rho^2, \label{E4} \\
  {\cal E}^{(0)}[\vec\phi] &= &\alpha V(\vec\phi), \label{E0}
\end{eqnarray}
where $\gamma$ and $\alpha$ are positive constants, and where $V(\vec\phi)$
is some suitable function of $\vec\phi$ (not involving its derivatives).
We regard the normalized energy as dimensionless.  The constants 
$\gamma$ and $\alpha$ have dimensions of length$^2$ and length$^{-2}$
respectively; the dimensionless combination $\alpha\gamma$ determines the
energy of a soliton, while its size is of order $(\gamma/\alpha)^{1/4}$.
In flat space, this latter quantity simply sets the length scale; but
on the sphere $M$, there already is a length scale $L$. So one has a
dimensionless parameter $\beta = L(\alpha/\gamma)^{1/4}$, which is
the ratio between the size of space and the size of the soliton.
From now on, we shall set $\gamma=1$.

The energy of any configuration of nonzero degree satisfies a generalized
Bogomol'nyi bound \cite{IRPZ92}.  First, $E^{(2)}$ satisfies the standard
O(3) sigma-model bound $E^{(2)}\geq 1$.  To get a bound on $E^{(4)}+E^{(0)}$,
write ${\cal E}^{(0)} = G^2/2$ and observe that
\begin{eqnarray*}
 E^{(4)}+E^{(0)} &= &\frac{1}{8\pi|k|}\int_M
             \bigl[(h^{-1}\rho/2-G)^2 + h^{-1}\rho\, G\bigr]\,\omega \\
                 &\geq &\frac{1}{8\pi|k|}\int_M\rho\, G\,dx^1\,dx^2 \\
                 &= &\frac{1}{4\pi}\int_{S^2}
                   G(\vec\phi)\,\,\vec\phi\cdot d\vec\phi\times d\vec\phi\,;
\end{eqnarray*}
the last equality follows from the degree theorem
$\int\phi^*(G\Omega)=k\int G\Omega$, with
$\Omega=\vec\phi\cdot d\vec\phi\times d\vec\phi$ being the area element
on the target sphere $S^2$.  If $G$ depends only on the third component
$\phi^3$ of $\vec\phi$, which is the
case for all the examples that we consider, then the bound on the total
energy takes the form
\begin{eqnarray}
 E_k &\geq &1 + \frac{1}{2}\int_{-1}^1 G(\phi^3)\,d\phi^3 \nonumber \\
     &= &1 + \sqrt{\alpha/2}\int_{-1}^1
          \sqrt{V(\phi^3)}\,d\phi^3. \label{Bog}
\end{eqnarray}

In this paper, we study rotationally-symmetric configurations.  Let us use
polar coordinates $(\theta,\varphi)$ on $M$, so that
$ds^2=L^2(d\theta^2 + \sin^2\theta\,d\varphi^2)$ and $h=L^2\sin\theta$.
We impose rotational symmetry by taking $\vec\phi$ to have the form
\begin{equation} \label{rot-sym}
 \vec\phi = \bigl(\sin(f)\cos(k\varphi),\sin(f)
                 \sin(k\varphi),\cos(f)\bigr),
\end{equation}
where $k$ is an integer and $f=f(\theta)$.  Assuming that this profile
function $f(\theta)$ satisfies the boundary conditions $f(0)=\pi$
and $f(\pi)=0$, the configuration (\ref{rot-sym}) has winding number $k$.
Its normalized energy is
\begin{equation} \label{rot-en}
 E_k = \frac{1}{4|k|}\int_0^{\pi}\biggl[(f')^2
        + k^2\biggl(\frac{\sin f}{\sin\theta}\biggr)^2
        + \biggl(\frac{kf'}{L}\frac{\sin f}{\sin\theta}\biggr)^2
        + 2L^2{\cal E}^{(0)}(\cos f) \biggr] \sin\theta\,d\theta,
\end{equation}
where $f'=df/d\theta$ and where we are (as mentioned above) taking
${\cal E}^{(0)}$ to be a function of $\phi^3=\cos f$ only.

We are interested in the stationary points of the functional (\ref{rot-en}),
and we find these numerically (with an accuracy better than $0.1\%$).  Two
independent methods were used: first, a direct minimization of (\ref{rot-en})
by a conjugate-gradient method; and secondly, solving the corresponding
Euler-Lagrange equations.

One particular `trial' profile which is of interest is the one corresponding
to a rotationally-symmetric conformal map ({\it cf} \cite{Man87}, \cite{SB98},
\cite{Kr00}).  Such a profile can be used for any value of $k$, but we shall
restrict here to the case $k=1$.  The profile function $f(\theta)$ is then
given by
\begin{equation} \label{hol-profile}
 \cot(f/2) = \lambda\tan(\theta/2),
\end{equation}
where $\lambda$ is a positive constant.  Notice that if $\lambda=1$,
then $f(\theta)=\pi-\theta$, and this corresponds to the identity map
from $M\cong S^2$ to $S^2$.  This identity map is only a solution of the
equations
of motion if ${\cal E}^{(0)}$ is constant ({\it ie} not depending on
$\vec\phi$).  Another way of putting this is that if (and only if)
${\cal E}^{(0)}$ is constant, then Skyrmions spread out to fill the whole
of space homogeneously, in the sense that the energy density ${\cal E}$
is constant.  The configuration (\ref{hol-profile})
has $E^{(2)}(\lambda) = 1$ and
$E^{(4)}(\lambda) = (\lambda^4+\lambda^2+1)/(6L^2\lambda^2)$.
Notice that $E^{(4)}(\lambda)$ has its minimum at $\lambda=1$; but the
location (and existence) of a minimum in the total energy $E(\lambda)$
depends on the choice of the potential term
$E^{(0)}$.  The question is: for a given $L$ and choice of $E^{(0)}$,
what are the minima of $E(\lambda)$, and in particular are they at
$\lambda=1$ or at $\lambda\neq1$?

We now wish to define the two quantities $\Phi$ and $\Psi$ mentioned in
the introduction.  The first of these is
\begin{equation} \label{Phi}
   \Phi = \langle (\phi^3)^2 \rangle  -\frac{1}{3}
       = \frac{1}{4\pi L^2} \int_M (\phi^3)^2 \,\omega\, -\frac{1}{3}\, .
\end{equation}
This is zero for the homogeneous configuration $f(\theta)=\pi-\theta$,
and so gives an indication of the deviation from homogeneity.
The second quantity is
\begin{equation} \label{Psi}
  \Psi = \langle \phi^3 \rangle^2
       = \biggl[\frac{1}{4\pi L^2} \int_M \phi^3 \,\omega\biggr]^2 \, .
\end{equation}
This is zero for configurations which have the reflection symmetry
$\phi^3\mapsto-\phi^3$, but nonzero if the Skyrmion is localized
around the point $\theta=0$ or $\theta=\pi$.


\section{Two asymmetric examples}

In this section, we give a brief discussion of two systems, arising from
two possible choices of ${\cal E}^{(0)}$.  Each is `asymmetric', in the
sense that neither posesses the symmetry $\phi^3\mapsto-\phi^3$.
So a Skyrmion in these systems will never have $\Psi=0$;  the field
prefers to be near $\vec\phi=(0,0,1)$ rather than $\vec\phi=(0,0,-1)$.
The first example (for corresponding
flat-space studies see \cite{LPZ90}, \cite{Sut91}) is motivated by
the question: can one saturate the Bogomol'nyi bound (\ref{Bog})?
We shall not give a complete analysis of this question here; but in the $k=1$
case, it is easy to derive the following fact: 
the configuration (\ref{hol-profile}) saturates the Bogomol'nyi bound
if and only if ${\cal E}^{(0)}$ is given by
\begin{equation} \label{hol-E0}
 {\cal E}^{(0)} = \frac{1}{2}
    \biggl[ \frac{(\lambda^2+1)-(\lambda^2-1)\phi^3}{2L\lambda}\biggr]^4.
\end{equation}
(Notice that the $\lambda=1$ case is degenerate: ${\cal E}^{(0)}$ is
then constant.)   For this system, the bound is
\begin{equation} \label{hol-bog}
 E \geq 1 + (\lambda^4+\lambda^2+1)/(3L^2\lambda^2).
\end{equation}
If, for example, we choose $\lambda = 2L$, then the configuration
(\ref{hol-profile}) is a solution with energy
\begin{equation} \label{hol-energy}
 E = \frac{28L^4+4L^2+1}{12L^4} = \frac{7}{3} + O(L^{-2});
\end{equation}
in the limit $L\to\infty$, this corresponds to \cite{LPZ90}, \cite{Sut91}
with their parameters $\theta_1$ and $\theta_2$ set equal to $0.5$ and $1$
respectively (note also that their expression for energy $E$ is $2\pi$
times ours).   The energy (\ref{hol-energy}) is a monotonic-decreasing
function of $L$; this remains true for variations of (\ref{hol-E0})
such as ${\cal E}^{(0)} = (1-\phi^3)^4$, and for $k>1$.

In this system, there is a repulsive force between solitons \cite{LPZ90},
\cite{Sut91}, \cite{ESZ00}.  Consequently, one expects that the
minimum-energy configurations for $k>1$ will not be rotationally-symmetric
(or, equivalently, that all rotationally-symmetric $k>1$ configurations
are unstable).  So investigation of the $k>1$ sectors requires a full
two-dimensional study.

The comments of the previous paragraph apply equally well to our other
example in this section, which was the one studied in \cite{PMTZ94},
\cite{PSZ95a}, \cite{PSZ95b} (on flat space) and \cite{SB98} (on the sphere).
For this system, we have $V(\phi^3)=1-\phi^3$.  The Bogomol'nyi bound
(\ref{Bog}) becomes $E_k\geq1+4\sqrt{\alpha}/3$ (note that this is stronger
than the bound given in \cite{SB98}).  Here we give a brief
description of the $k=1$ case, taking $\alpha=0.1$. 
For the $\lambda$-approximation (\ref{hol-profile}), we have
\begin{equation} \label{OBS}
 E^{(0)}(\lambda) = \frac{2L^2\alpha}{1-\lambda^2}
   \biggl[ 1 + \frac{\lambda^2\log(\lambda^2)}{1-\lambda^2} \biggr],
\end{equation}
which is a decreasing function of $\lambda$.
The total energy $E(\lambda)$ has a minimum at $\lambda=\lambda_{{\rm min}}$,
with $\lambda_{{\rm min}} > 1$ depending on the parameter
$\beta=L\alpha^{1/4}$.   In figure 1,
we plot the energy $E(\lambda_{{\rm min}})$, the actual Skyrmion energy
$E=E_1$ (the minimum of (\ref{rot-en})), and the quantities 
\begin{figure}[hb]
\begin{center}
\subfigure[Energies $E$ and $E(\lambda_{{\rm min}})$]{
\includegraphics[scale=0.3]{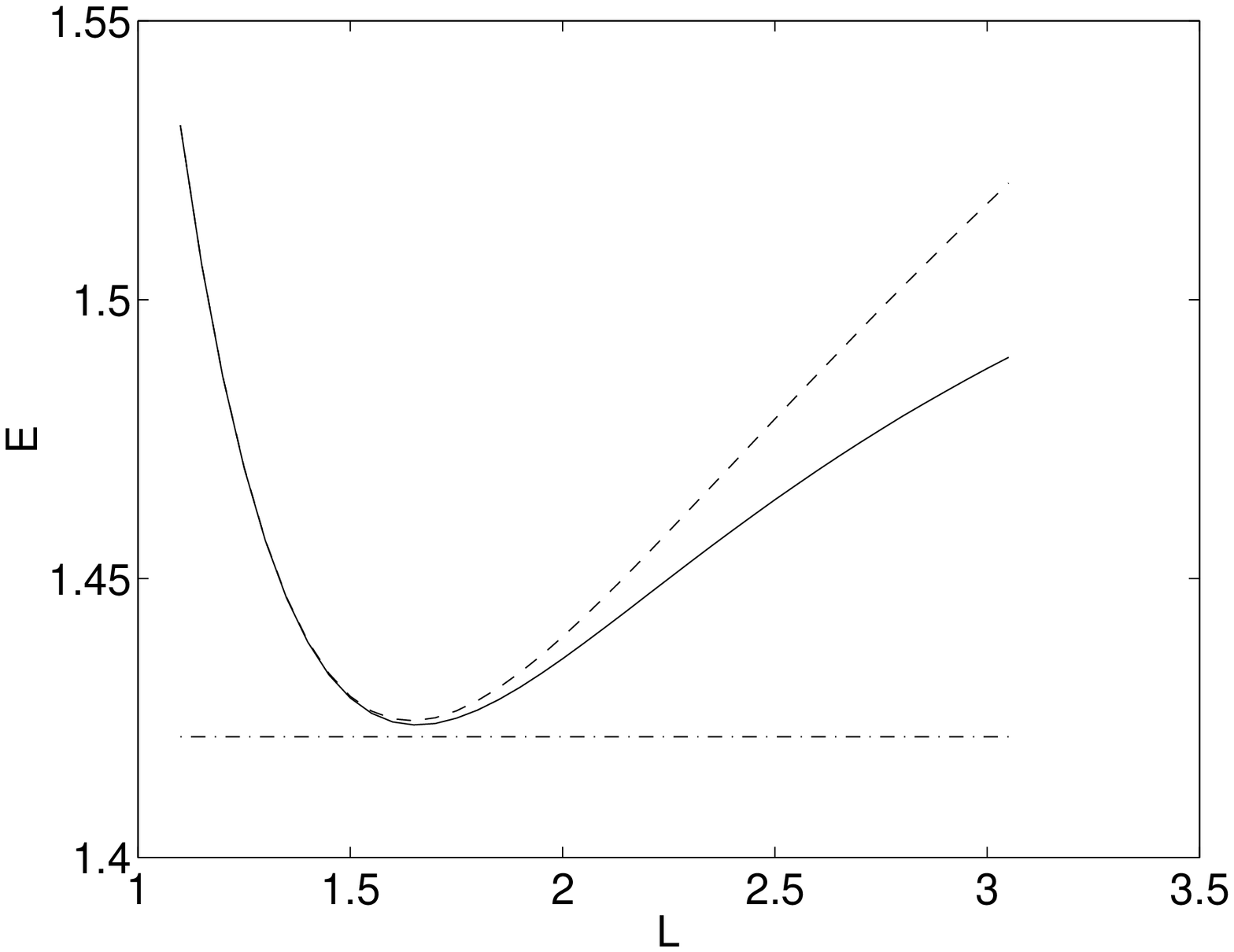}
}
\quad
\subfigure[$\Psi$]{
\includegraphics[scale=0.3]{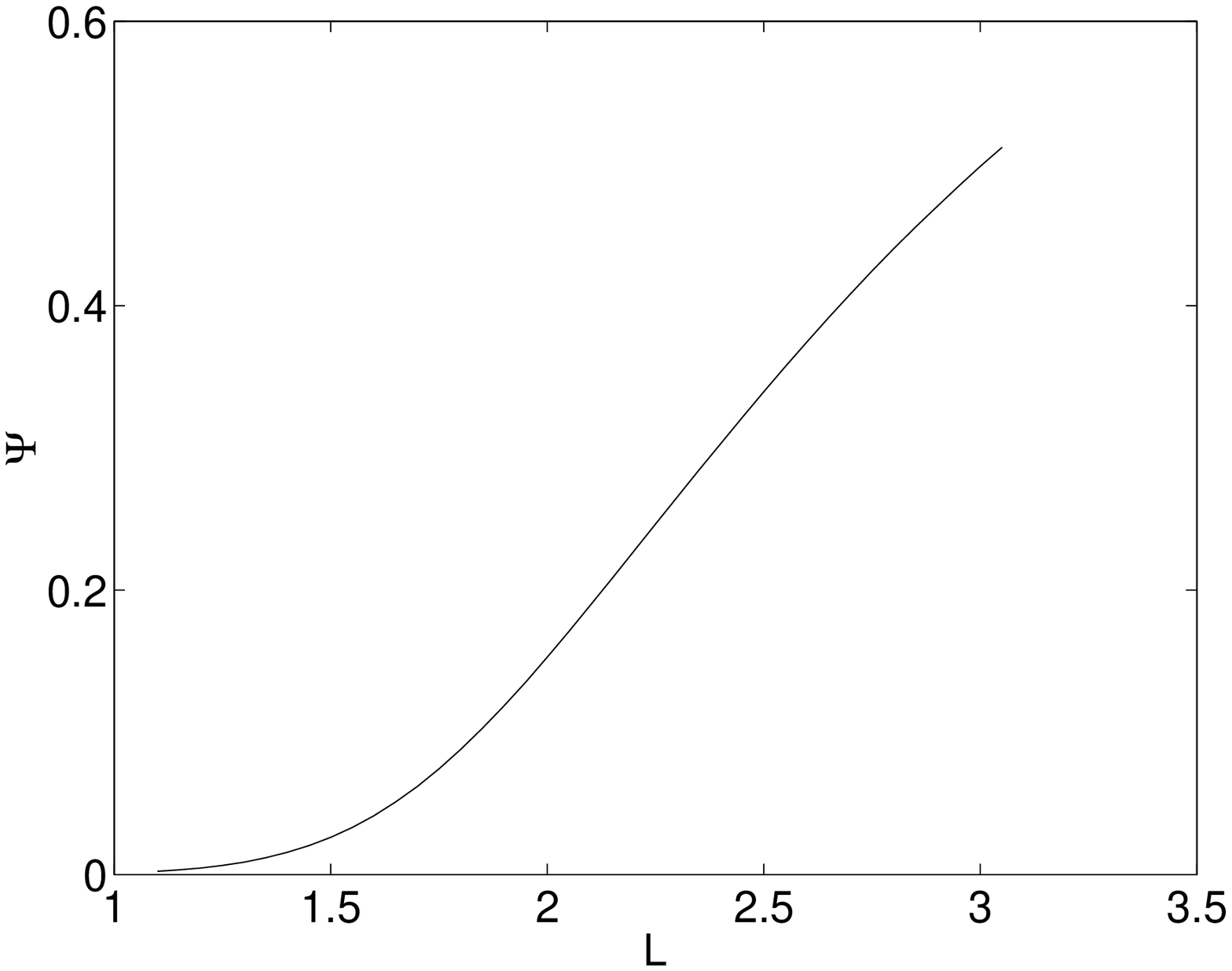}
}
\quad
\subfigure[$\Phi$]{
\includegraphics[scale=0.3]{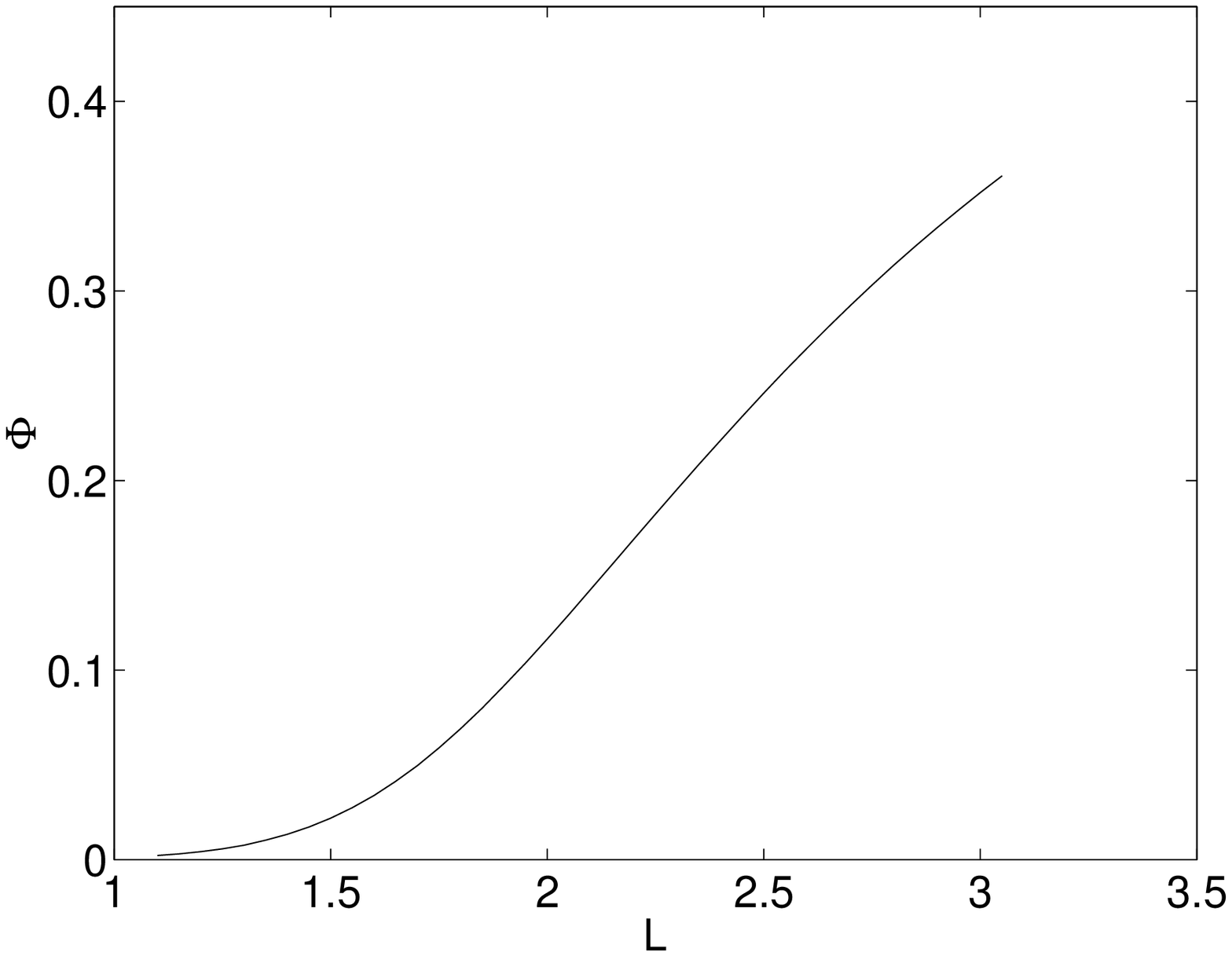}
}
\quad
\subfigure[$\lambda_{{\rm min}}$]{
\includegraphics[scale=0.3]{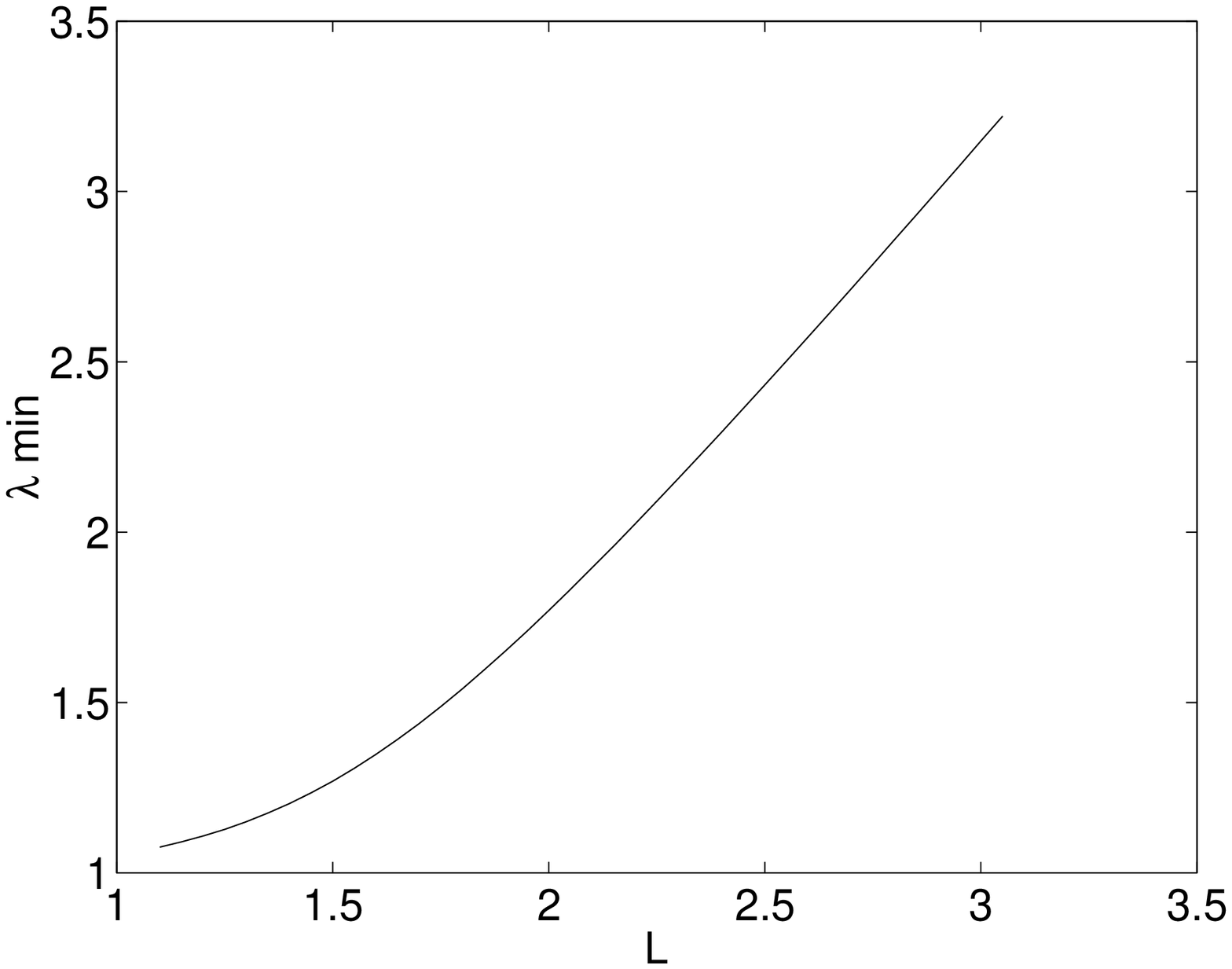}
}
\caption{Properties of a single Skyrmion, for the system with $V=1-\phi^3$,
    as a function of $L$. In (a), the dot-dashed line is the Bogomol'nyi bound,
    the solid curve is the actual energy, and the dashed curve is the
    best $\lambda$-approximation.\label{fig1}}
\end{center}
\end{figure}
$\Psi$, $\Phi$ and $\lambda_{{\rm min}}$, all as functions of $L$.
Notice that the lowest value of $E$ (when $L=1.65$ and $\beta=0.93$) is
$1.424$, which less than $0.2\%$ greater than the Bogomol'nyi bound $1.4216$.
At this value of $L$, we have $\lambda_{{\rm min}}=1.39$ and
$E(\lambda_{{\rm min}})=1.425$.   As $L\to\infty$, the energy of
the Skyrmion tends to its flat-space value of $1.549$ \cite{We99}.

As was noted in \cite{SB98}, there is a regime ($\beta<1$) where the Skyrmion
is approximately spread out over $M$ and well-approximated by the
configuration (\ref{hol-profile}), and one ($\beta>1$) where the Skyrmion
is localized and not well-approximated by (\ref{hol-profile}).  But the
transition between these two regimes is not sharply-defined, and is not a
phase transistion in the usual sense.   This is partly due to the
asymmetry of the system.  In our next examples, we shall see a sharper
transition.


\section{Two symmetric examples}

In this section, we study the systems which are defined, respectively, by
$V(\phi^3) = 1-(\phi^3)^2$ and $V(\phi^3)= (\phi^3)^2[1-(\phi^3)^2]$.
In these systems, Skyrmions attract one another \cite{We99}, \cite{ESZ00};
and consequently the solutions with $k>1$ have rotational symmetry.
Since these potentials have the symmetry $\phi^3\mapsto-\phi^3$, one
expects (in the rotationally-symmetric case, and for large $L$) that
there will be a stable Skyrmion solution localized at a point (either 
$\theta=0$ or $\theta=\pi$); and also that there exists a solution which is
symmetric under $\theta\mapsto\pi-\theta$  (and which may or may not be
stable).  For small $L$, however, only the symmetric solution exists.

First take $V(\phi^3) = 1-(\phi^3)^2$, with the value $\alpha=0.08873$
(this corresponds to the value used in the flat-space study of \cite{We99},
and allows a quantitative comparison with its results). The Bogomol'nyi bound
(\ref{Bog}) is $E \geq 1+\pi\sqrt{\alpha/8}=1.3309$.  Let us look first
at the $k=1$ sector. For the $\lambda$-approximation (\ref{hol-profile}),
we have
\begin{equation} \label{NBS1}
 E^{(0)}(\lambda) = \frac{4\alpha L^2\lambda^2}{(1-\lambda^2)^2}
   \biggl[ -2 - \frac{1+\lambda^2}{1-\lambda^2}\log(\lambda^2) \biggr],
\end{equation}
which is a positive function with a maximum at $\lambda=1$ (recall that
$E^{(4)}(\lambda)$ has a minimum at $\lambda=1$).  Notice that
$E(\lambda^{-1}) = E(\lambda)$; this corresponds to the
$\phi^3\mapsto-\phi^3$ symmetry of $E[\vec\phi]$ in this case.
So $\lambda=1$ gives either a minimum or a local maximum of $E(\lambda)$,
depending on the value of $L$.
A simple calculation reveals that if $\beta^4 \leq 5/2$, then
\begin{equation} \label{NBS2}
 E(1) = 1+\frac{1}{2L^2}+\frac{2L^2\alpha}{3}
\end{equation}
is a minimum of $E(\lambda)$; while if $\beta^4 > 5/2$, then there is a local
maximum at $\lambda=1$, and the minima are at $\lambda_{{\rm min}}>1$
and $1/\lambda_{{\rm min}}$.  Consequently, if we restrict to the special
profiles (\ref{hol-profile}), then there is a sharp phase transition at
$\beta^4 = 5/2$ between a homogeneous phase ($\lambda=1$) and one where
the Skyrmion localizes at a point (in this case, one of the poles) of the
sphere $M$.  Notice also that the lowest value of $E(1)$, namely
$E=1+\sqrt{4\alpha/3}=1.344$,
is attained when $\beta^4 = 3/4 \Leftrightarrow L = 1.705$.

This is just an approximation; the true Skyrmion energy is strictly
less than $E(\lambda)$.  However, there is indeed a minimum at $L = 1.705$
(with $E=1.341$, only $0.2\%$ less than the $\lambda$-approximation);
and there is a transition at
$\beta^4 = 5/2 \Leftrightarrow L = 2.304$, as one can see from figure 2.
\begin{figure}[htb]
\begin{center}
\subfigure[Energies for $k=1$]{
\includegraphics[scale=0.3]{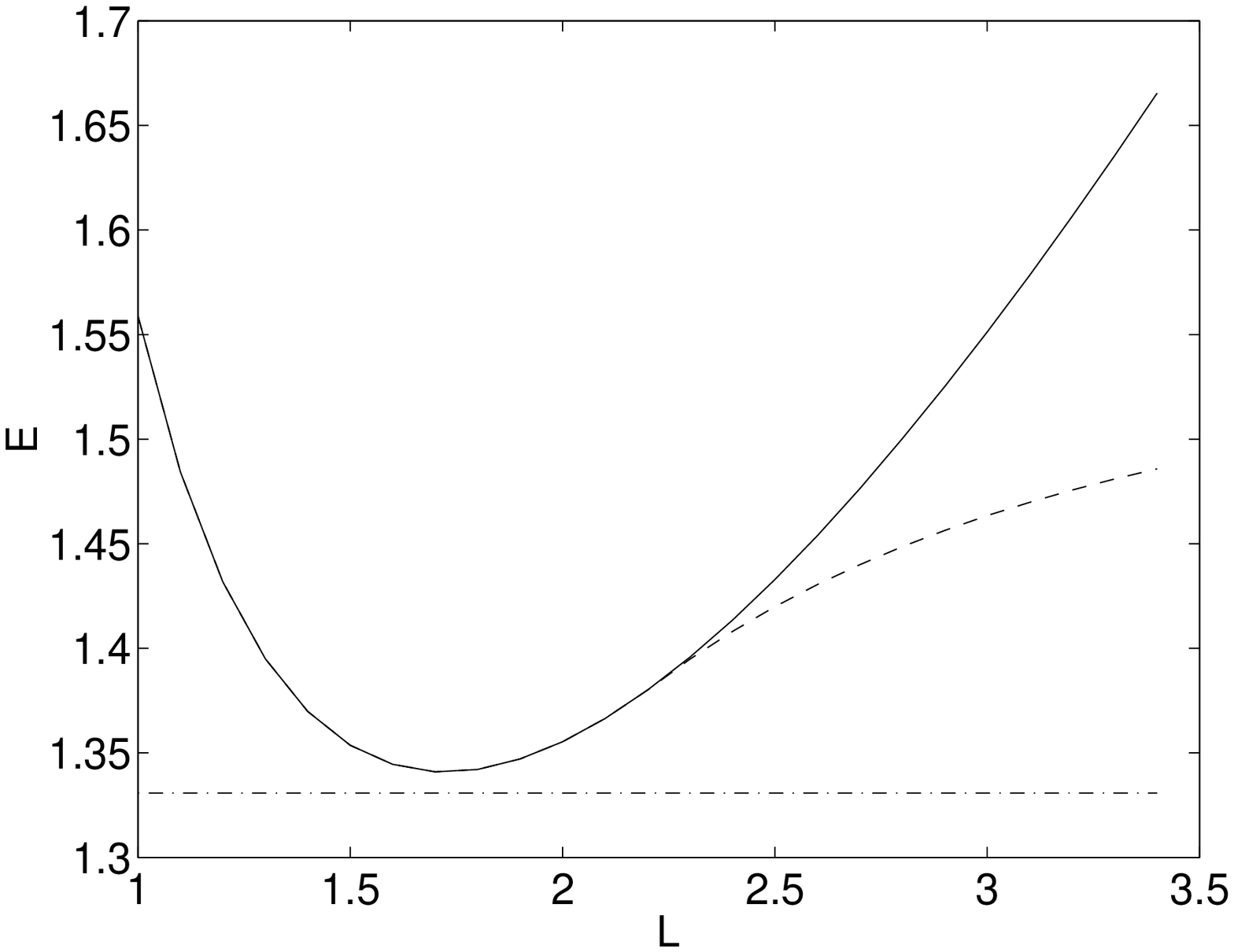}
}
\quad
\subfigure[$\Psi$]{
\includegraphics[scale=0.3]{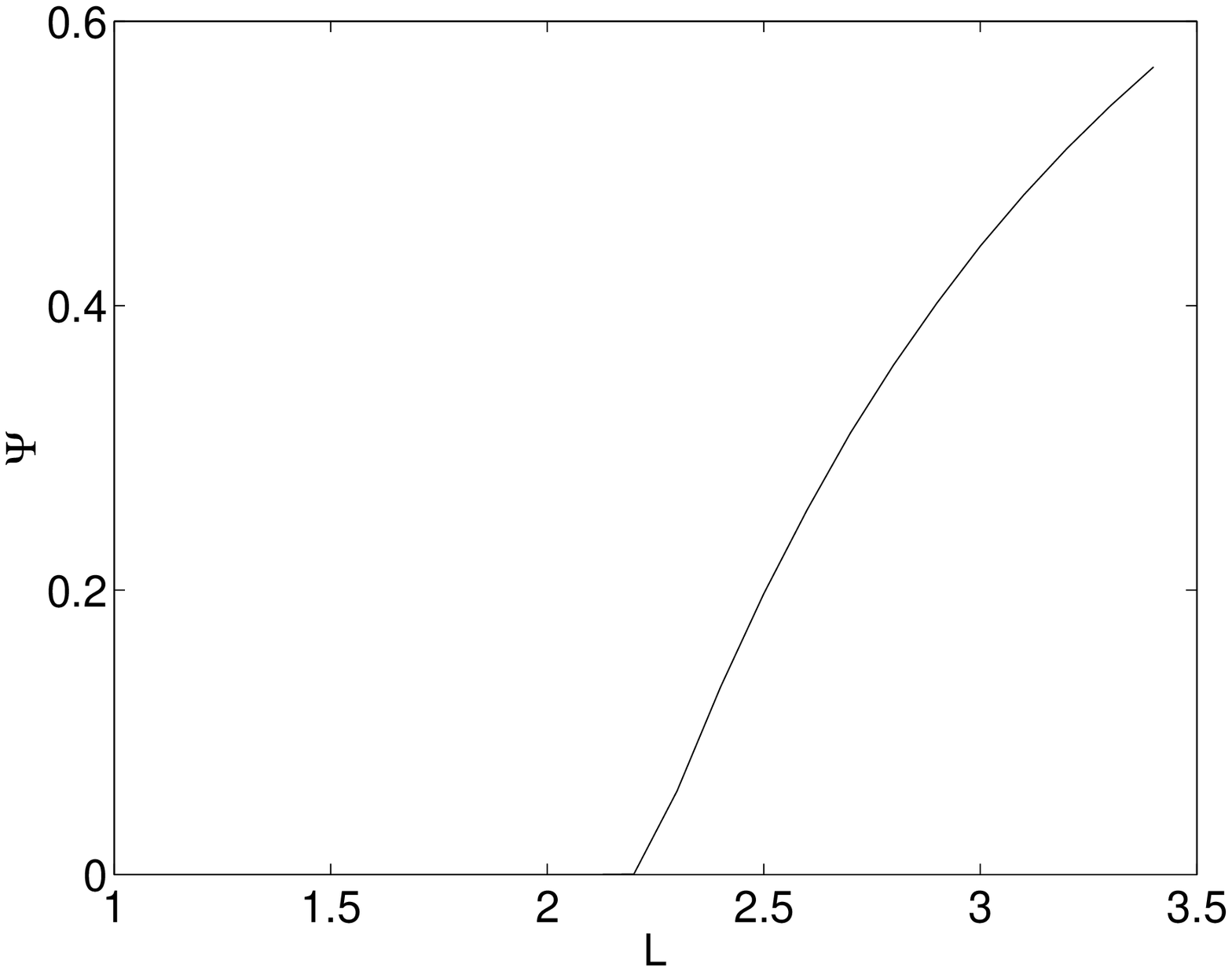}
}
\quad
\subfigure[$\Phi$]{
\includegraphics[scale=0.3]{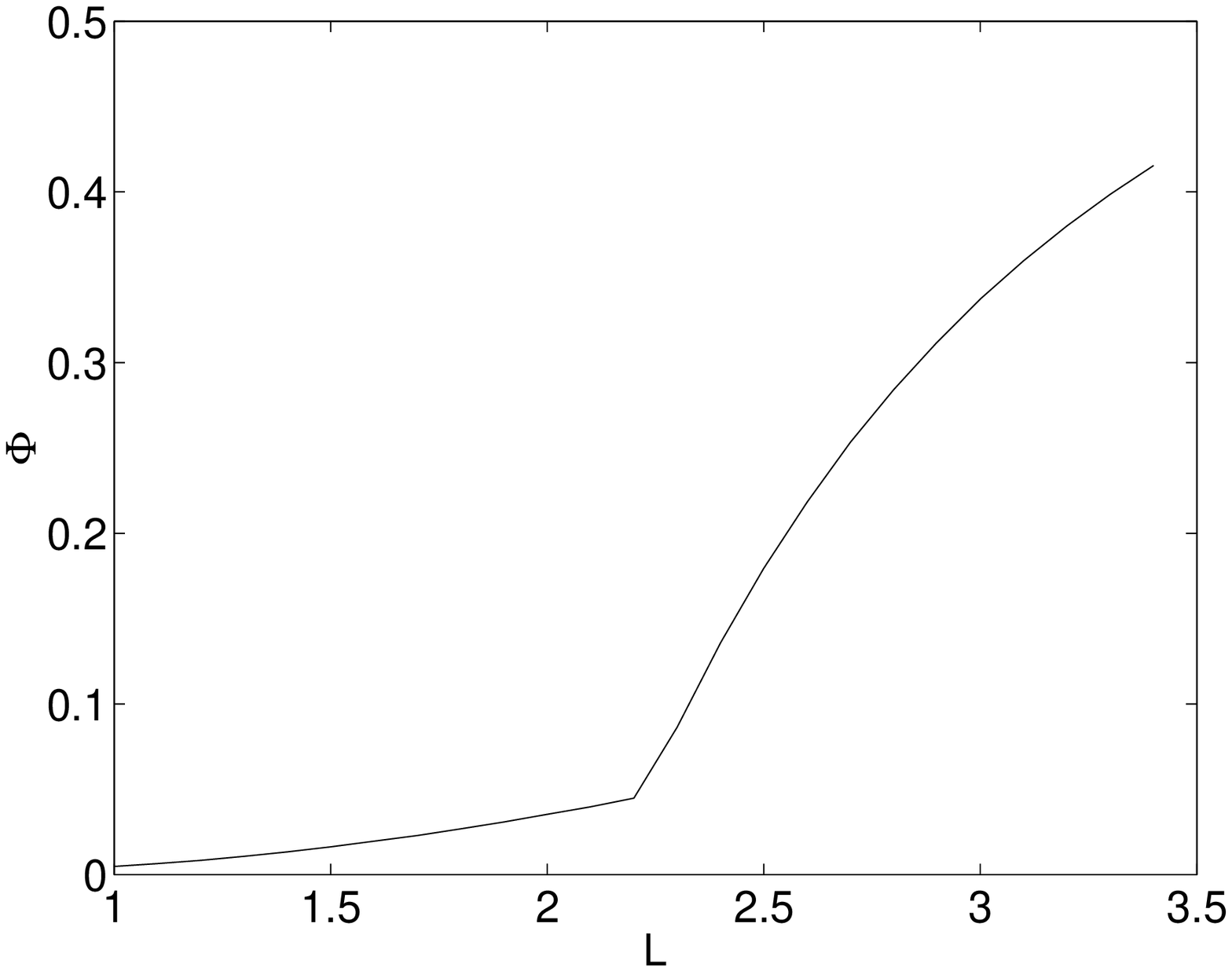}
}
\quad
\subfigure[Energies for $k=1,2,3$]{
\includegraphics[scale=0.3]{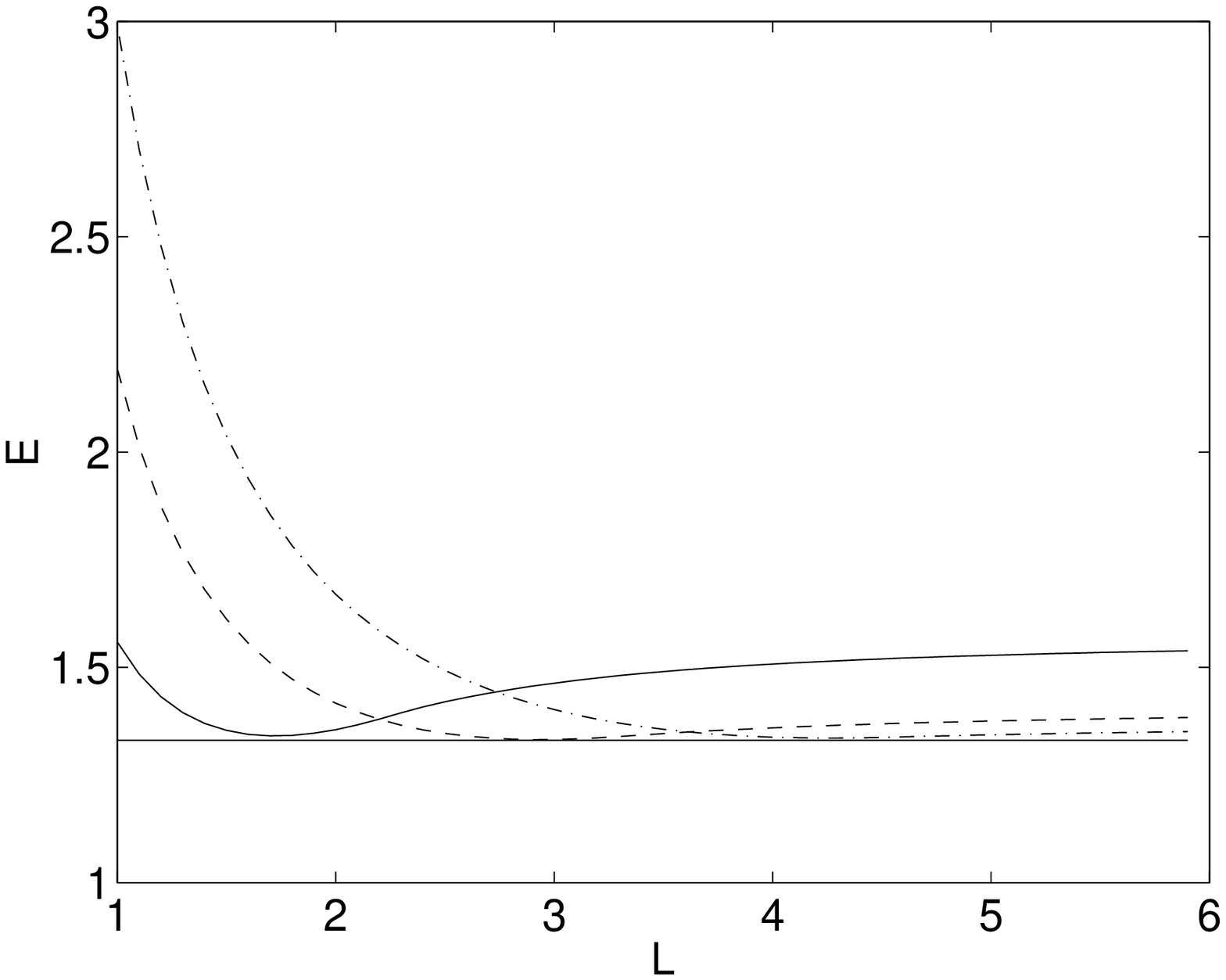}
}
\caption{Skyrmion properties for the system with
  $V=1-(\phi_3)^2$, as a function of $L$.
  In (a), the dot-dashed line is the Bogomol'nyi bound; the solid curve
  is the energy of the symmetric solution; and the dashed curve is
  the energy of the asymmetric solution.  Figures (b) and (c)
  plot the quantities $\Psi$ and $\Phi$ for the $k=1$ minimum.
  Figure (d) gives minimum energies for $k=1$ (solid curve), $k=2$
  (dashed curve) and $k=3$ (dot-dashed curve); the solid line is the
  Bogomol'nyi bound.\label{fig2}}
\end{center}
\end{figure}
Figure 2(a) depicts the energies of two static $k=1$ solutions, as functions
of $L$.  The solid curve is the energy of the `symmetric' solution, {\it ie}
the one satisfying $f(\pi-\theta)=\pi-f(\theta)$.  This is obtained by
starting with the configuration $f(\theta) = \pi-\theta$ and iterating
towards a stationary point; our procedures are able to converge to a
saddle-point (which this is for large $L$). The dashed curve is the energy
of an asymmetric solution, obtained by starting with a highly-asymmetric
configuration.  We see that for $\beta^4 < 5/2$, the two solutions coincide;
but for $\beta^4 > 5/2$, the former becomes unstable (numerical experiment
indicates that it is a saddle-point rather than a local minimum).  The
corresponding parameters $\Psi$ and $\Phi$ for the minimum-energy solution
are plotted in figures 2(b) and 2(c); note that when $\beta^4 < 5/2$,
we have $\Psi=0$ (the solution is symmetric), but $\Phi\neq0$ (it
is not homogeneous).  There is a second-order phase transition at
$\beta^4 = 5/2$.  In figure 2(d), the Skyrmion energies (minima of $E$)
are plotted for $k=1,2,3$.  It might be noted that $E_k$ gets very close
to the Bogomol'nyi bound, at $L=1.7$, $L=2.9$ and $L=4.3$ for $k=1,2,3$
respectively (only $0.1\%$ above for $k=2$).  As $L\to\infty$, we have
\cite{We99}
\begin{equation}
  E_1 \to1.564, \quad E_2 \to1.405, \quad E_3 \to1.371.
\end{equation}
%


Now let us turn to the system with $V(\phi^3)=(\phi^3)^2[1-(\phi^3)^2]$,
taking $\alpha=0.04$ (which is consistent with the choice made in
\cite{ESZ00}).  The Bogomol'nyi bound is $E \geq 1+\sqrt{2\alpha}/3=1.0943$.
In the $k=1$ sector, the $\lambda$-approximation (\ref{hol-profile}) has
\begin{equation} \label{NNBS}
 E^{(0)}(\lambda) = \frac{4L^2\alpha\lambda^2\bigl[
      (10+18\lambda^2-18\lambda^4-10\lambda^6) +
      (3+21\lambda^2+21\lambda^4+3\lambda^6)\log(\lambda^2)\bigr]}
      {3(\lambda^2-1)^5},
\end{equation}
which has a local minimum at $\lambda=1$, with $E^{(0)}(1)=2L^2\alpha/15$.
Consequently, $E(\lambda)$ always has a local minimum at $\lambda=1$.
Notice that for $L=[15/(4\alpha)]^{1/4}=3.11$
(where the true energy has its lowest value $1.1026$), the approximation
has energy $E(1) = 1+\sqrt{4\alpha/15} = 1.1033$. 

There are two values  $L_1$ and $L_2$ which are critical in the following
sense.  For $L\leq L_1$, $E(\lambda)$ has only one stationary point (the
minimum at $\lambda=1$).  For $L>L_1$, $E(\lambda)$ has three local minima:
a symmetric one ($\lambda=1$) and a degenerate asymmetric one ($\lambda\neq1$).
For $L_1<L<L_2$, the asymmetric minimum has higher energy than the symmetric
one: see figure 3(a), where $E$ is plotted as a function of
$\mu=(\lambda^2-1)/(\lambda^2+1)$, for $L=7.4$.  Finally, for $L>L_2$,
the asymmetric minimum has the lower energy ({\it cf.} figure 3(b),
for $L=8.4$); and the energy of the symmetric minimum tends to infinity
as $L\to\infty$.
\begin{figure}[htb]
\begin{center}
\subfigure[Energy $E(\mu)$ for $L=7.4$]{
\includegraphics[scale=0.3]{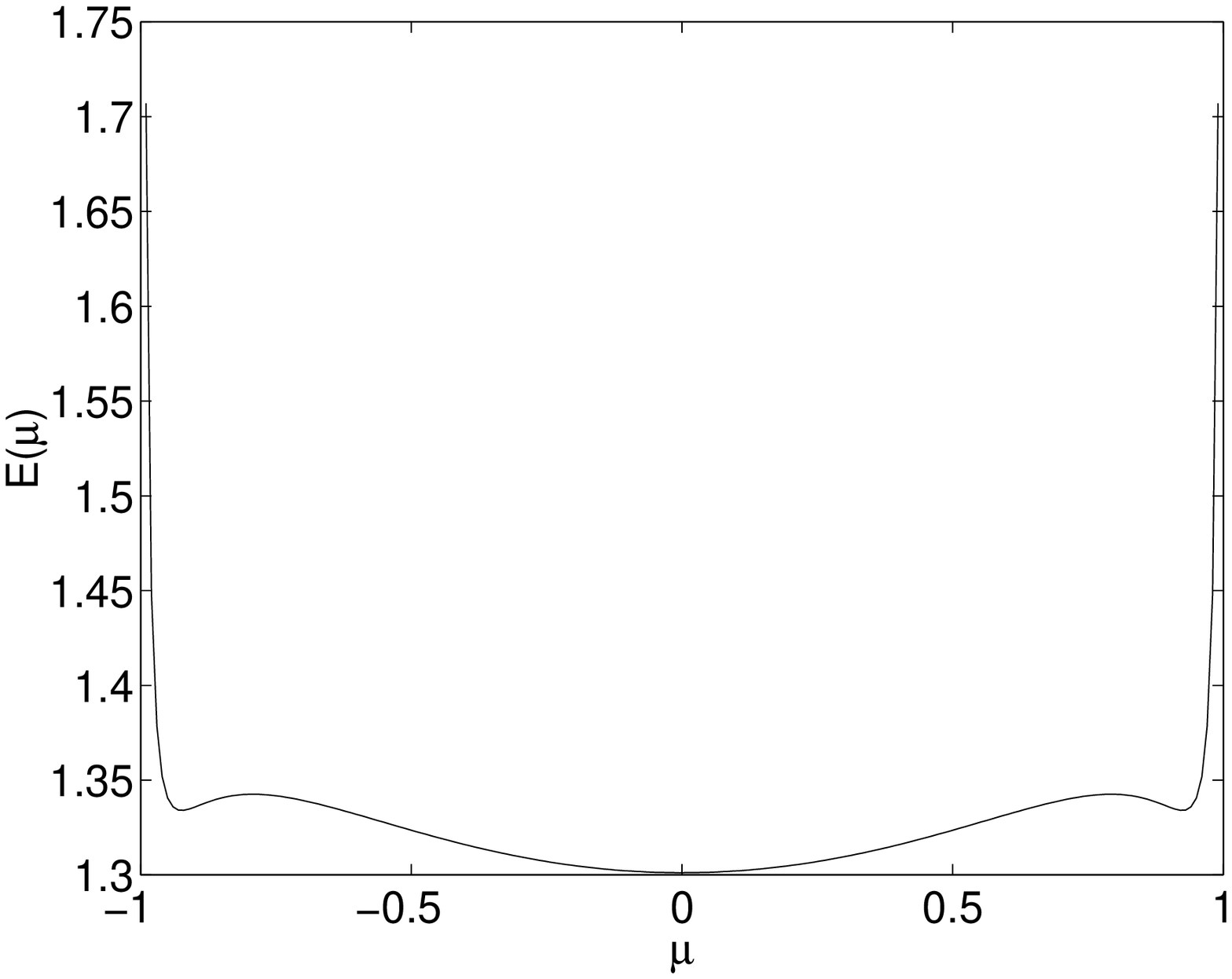}
}
\quad
\subfigure[Energy $E(\mu)$ for $L=8.4$]{
\includegraphics[scale=0.3]{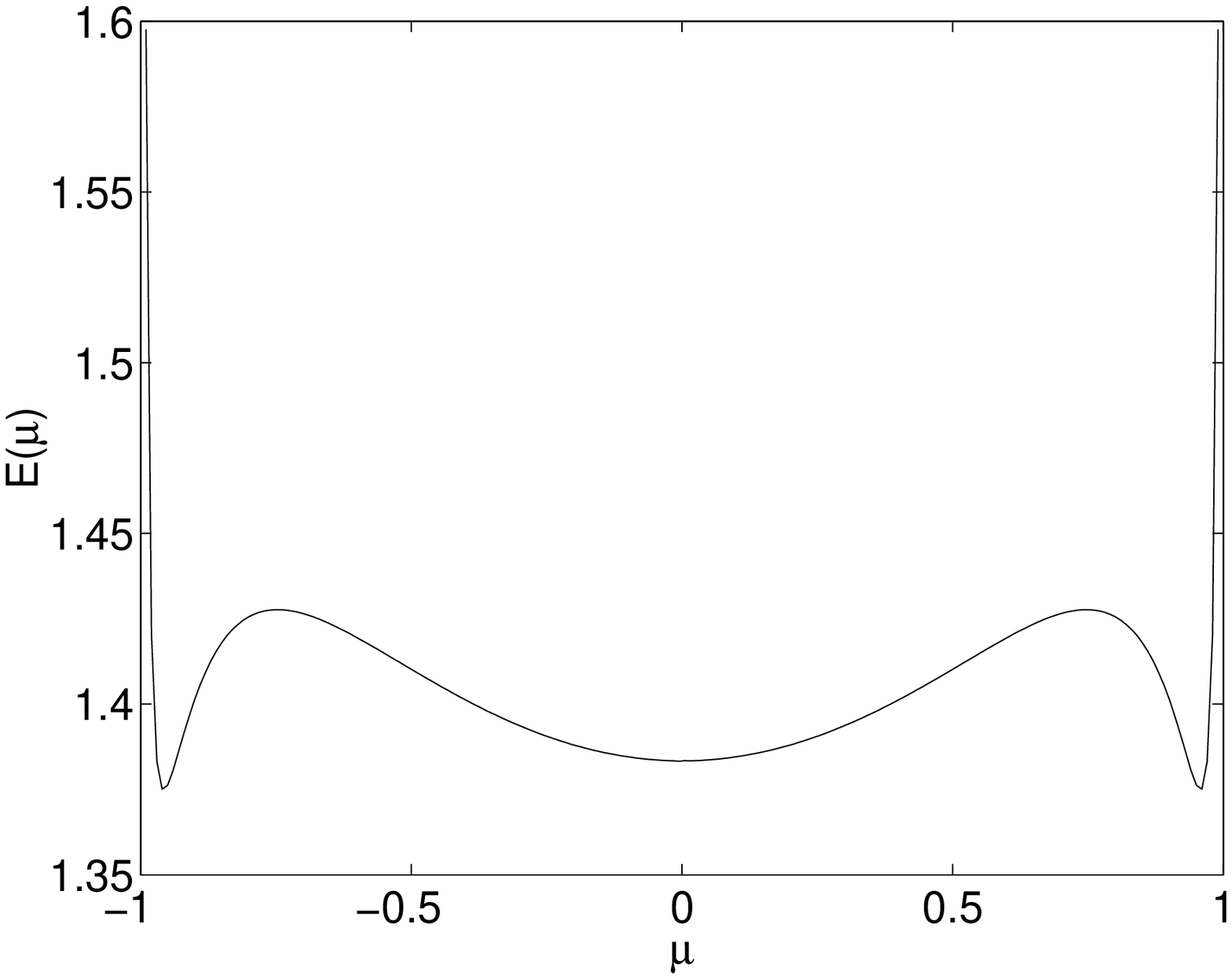}
}
\caption{Energy $E(\mu)$ of the approximation (\ref{hol-profile}),
  for the system with $V=(\phi_3)^2[1-(\phi_3)^2]$, where
  $\lambda^2=(1+\mu)/(1-\mu)$.  For $L=7.4$, the symmetric
  ($\mu=0$, $\lambda=1$) minimum has lower energy than the asymmetric
  minima.  For $L=8.4$, the asymmetric minima are lower.\label{fig3}}
\end{center}
\end{figure}

This approximate picture suggests that for $L$ large enough, there are
two stable Skyrmion solutions (symmetric and asymmetric); but that only
the latter survives in the limit as $L\to\infty$.  Our numerical investigation
shows that this is indeed the case.  Results from minimizing (\ref{rot-en})
are presented in figure 4.
\begin{figure}[htb]
\begin{center}
\subfigure[Energies for $k=1$]{
\includegraphics[scale=0.3]{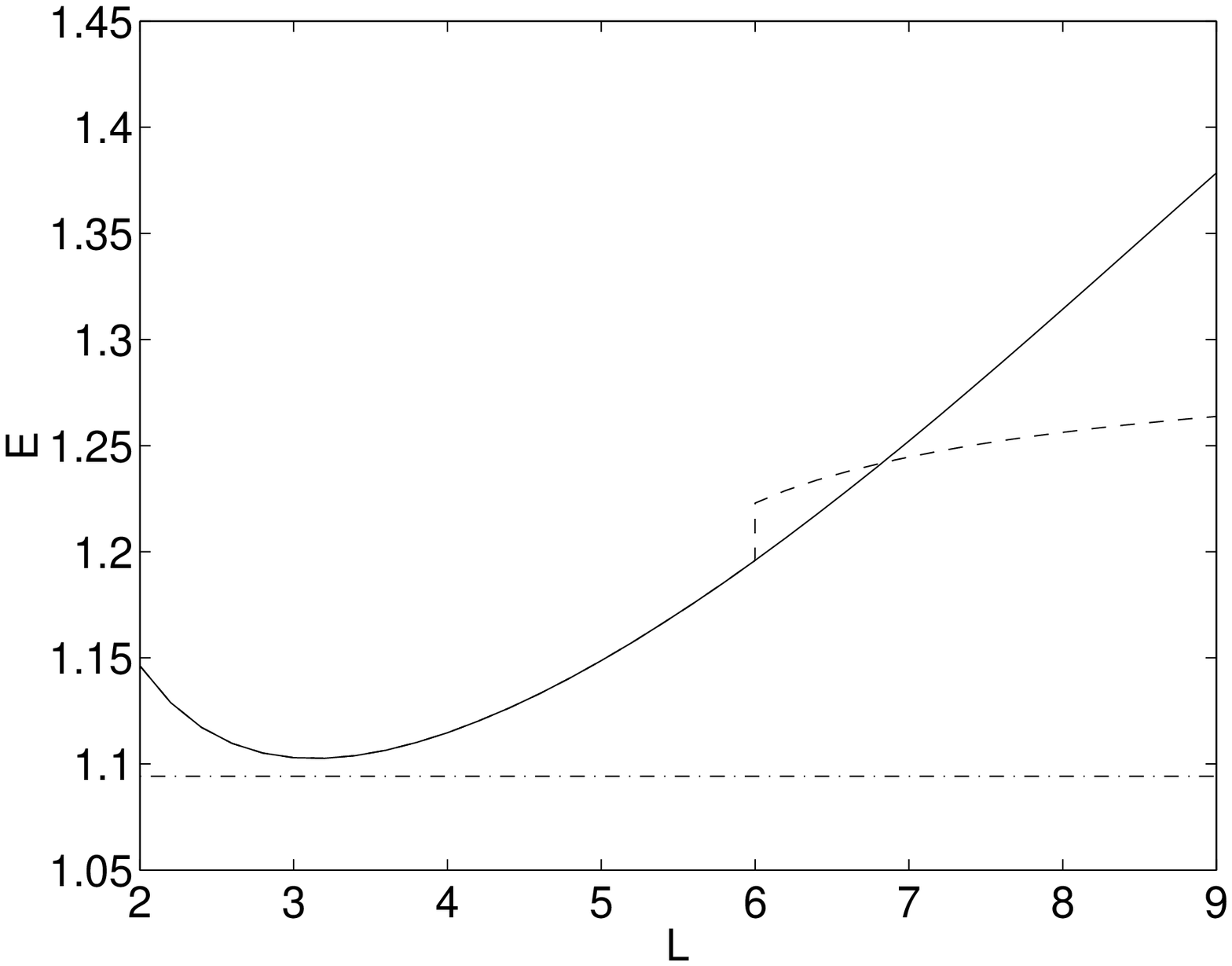}
}
\quad
\subfigure[$\Psi$]{
\includegraphics[scale=0.3]{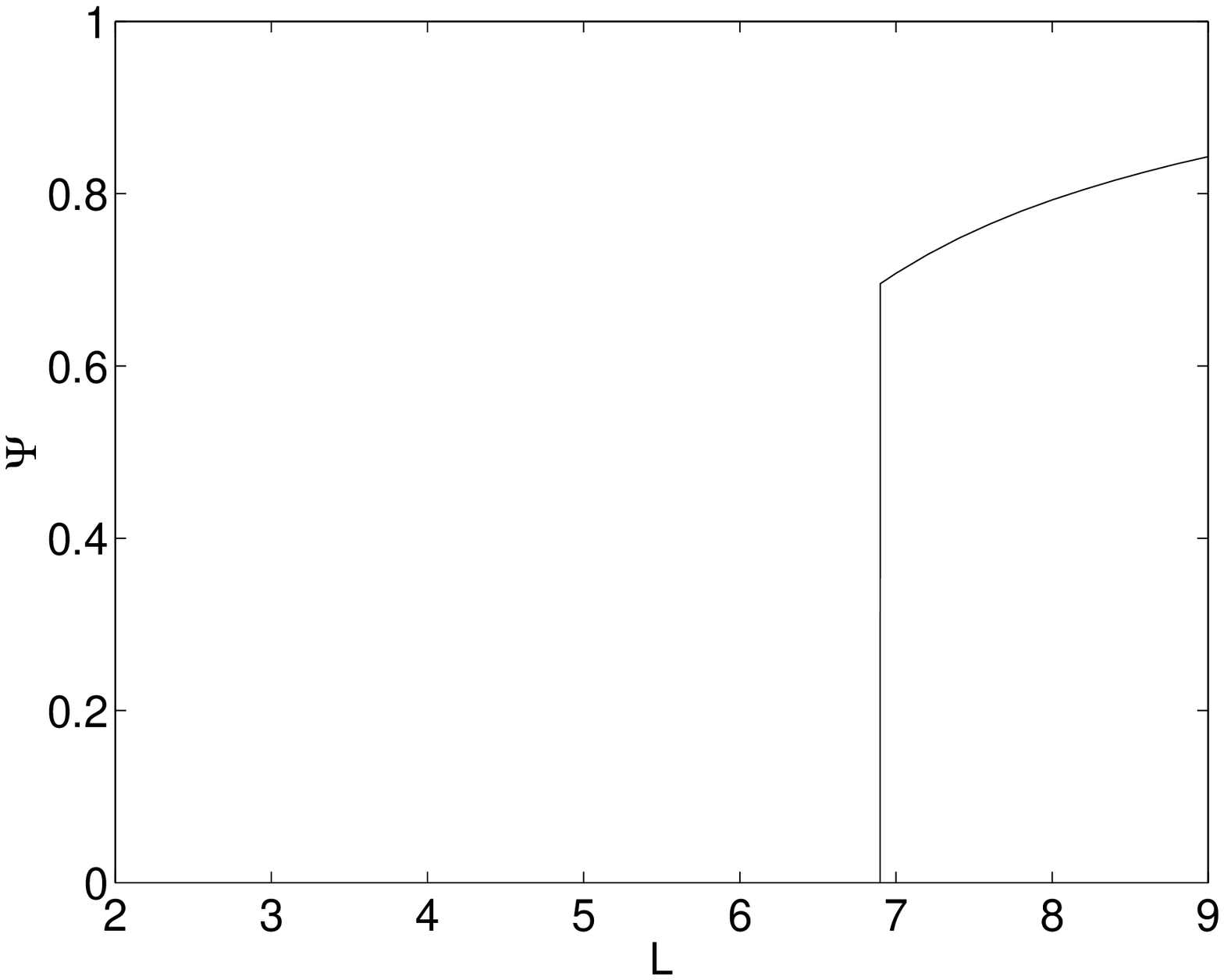}
}
\quad
\subfigure[$\Phi$]{
\includegraphics[scale=0.3]{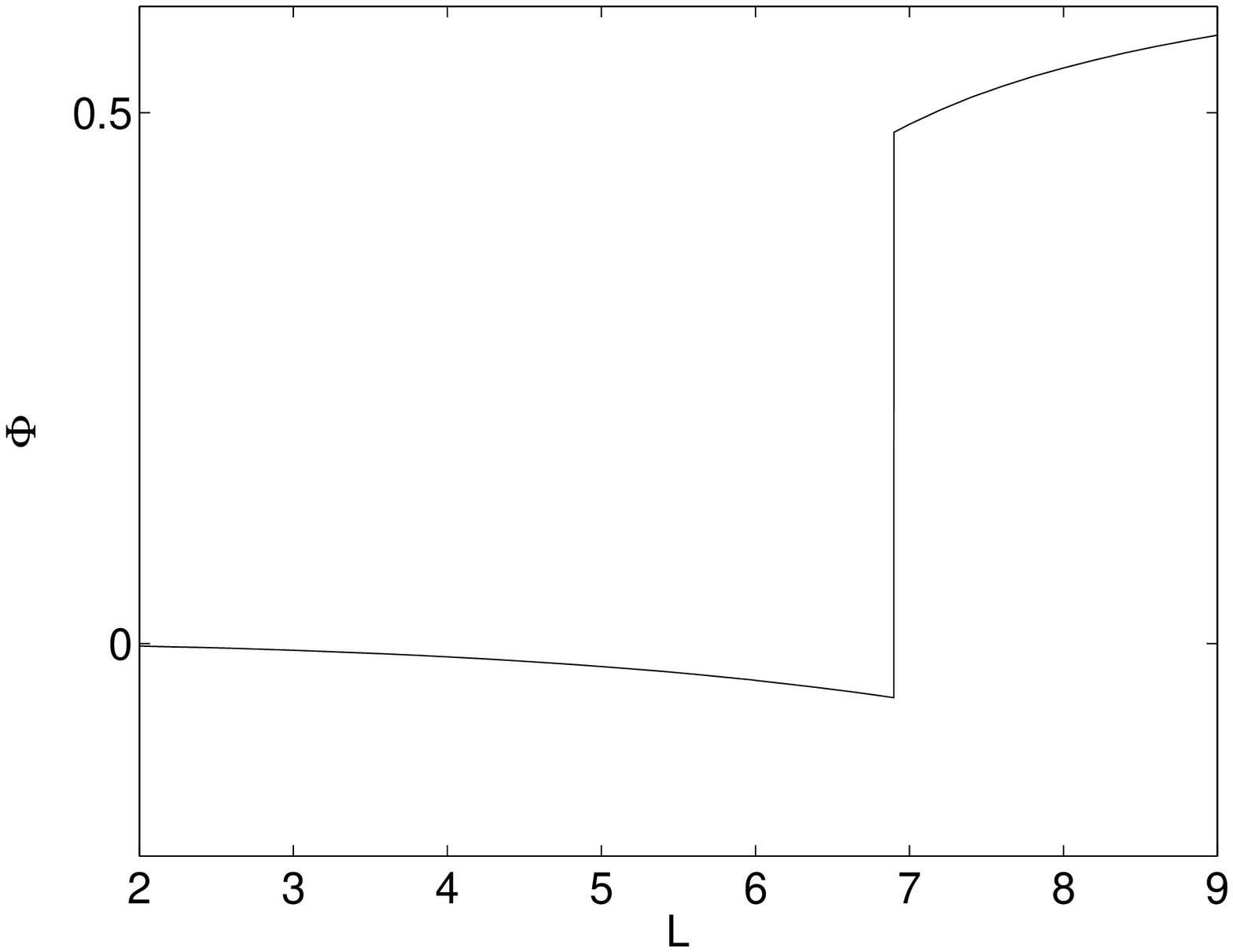}
}
\quad
\subfigure[Energies for $k=1,2,3$]{
\includegraphics[scale=0.3]{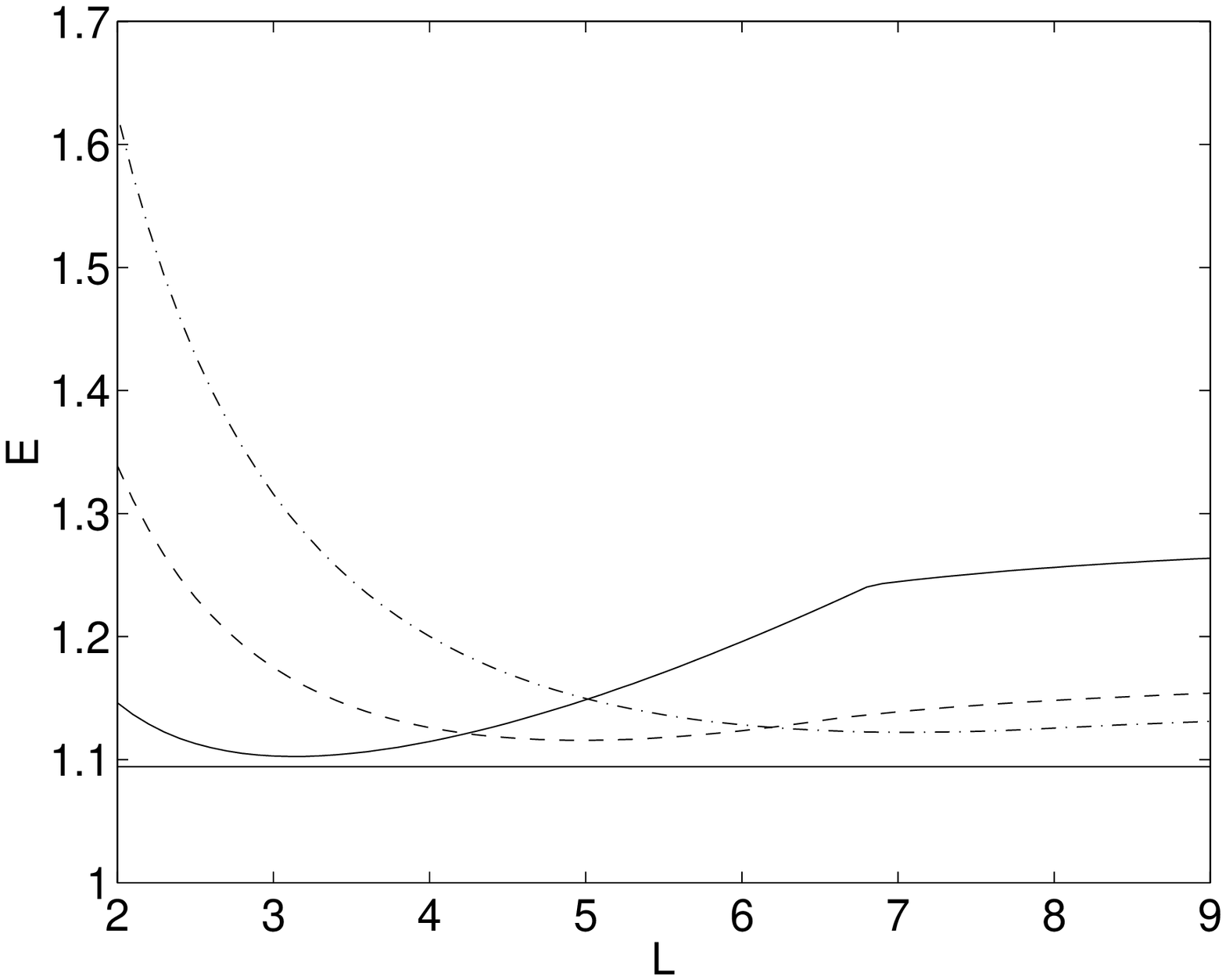}
}
\caption{Skyrmion properties for the system with
  $V=(\phi_3)^2[1-(\phi_3)^2]$, as a function of $L$.
  In (a), the dot-dashed line is the Bogomol'nyi bound; the solid curve
  is the energy of the symmetric solution; and the dashed curve is
  the energy of the asymmetric solution.  Figures (b) and (c)
  plot the quantities $\Psi$ and $\Phi$ for the minimal-energy $k=1$ solution.
  Figure (d) gives the energy for $k=1$ (solid curve), $k=2$
  (dashed curve) and $k=3$ (dot-dashed curve), for the minimal-energy
  solutions. \label{fig4}}
\end{center}
\end{figure}
In figure 4(a), the solid curve is the energy of the symmetric $k=1$ solution;
while the dashed curve is the energy of the asymmetric solution, which
exists only for $L\geq L_1\approx6$.  For $L_1<L<L_2\approx6.9$, the energy of
the asymmetric solution is greater than that of the symmetric solution.
Numerical evidence indicates that
each of these solutions is stable ({\it ie} is a local minimum of the energy),
at least in the rotationally-symmetric class.  The stability of the
symmetric solution, even for large $L$, is a consequence of the form of the
potential $V=(\phi^3)^2[1-(\phi^3)^2]$, which favours
$\phi^3=0 \Leftrightarrow f(\theta)=\pi/2$.  It also favours $\phi^3=\pm1$,
which stablizes the asymmetric solution.  Between these local minima
there will be a saddle-point solution, but we have not investigated this.

Figures 4(b) and 4(c) show the quantities $\Psi$ and $\Phi$ for the
minimal-energy $k=1$ solution; and in figure 4(d), the minimal energy
is plotted for $k=1,2,3$.   As $L\to\infty$, we have \cite{ESZ00}
\begin{equation}
  E_1 \to1.281, \quad E_2 \to1.166, \quad E_3 \to1.143.
\end{equation}
%

\section{Concluding Remarks}

We have studied rotationally-symmetric static solutions of the
Skyrme model on the two-sphere of radius $L$.  Even in the sector
with winding number $k$ equal to unity, the properties of the solutions,
and of the transition between the high-density (small $L$) and
low-density (large $L$) phases, depends crucially on the choice of
potential term.

It should be instructive to investigate the $k>1$ solution spaces
in more detail ({\it cf} \cite{Kr00}, \cite{ESZ00}, \cite{We99}).
For symmetric potentials (such as those of section 4),
this can be done within the rotationally-symmetric class; this is
the analogue of the rational-map ansatz \cite{Kr00} in three dimensions.
But for asymmetric potentials such as those of section 3, one expects
a different picture: the Skyrmions should separate and form some pattern
on $S^2$ as their optimal configuration.  Flat-space studies \cite{ESZ00}
indicate that this pattern will, in general, be rather complicated.

\bibliographystyle{plain} \bibliography{paper_ref}

\end{document}